\documentclass[letter]{aa} 

\usepackage{graphicx}
\usepackage{epstopdf}
\usepackage{deluxetable}
\usepackage{times,epsfig} 
\usepackage{mathrsfs}  
\usepackage{natbib}
\usepackage{amssymb}
\bibpunct{(}{)}{;}{a}{}{,} 
\usepackage{caption}

\usepackage[english]{babel}
\usepackage{longtable}
\usepackage{url}
\usepackage{hyperref}

\authorrunning{Taddia et al.}
\titlerunning{An observational study of SN~2008J.}
\begin{document}

\title{Supernova 2008J: early time observations of a heavily reddened SN~2002ic-like transient\thanks{Based on observations
 collected at the European Organisation for Astronomical Research in the Southern Hemisphere, Chile (ESO Programme 080.A--0516).}}

\author{F. Taddia\inst{1}
\and M.~D. Stritzinger\inst{2}
\and M.~M. Phillips\inst{3}
\and C.~R. Burns\inst{4}
\and E. Heinrich-Josties\inst{4}
\and N. Morrell\inst{3}
\and J. Sollerman\inst{1}
\and S. Valenti\inst{5}
\and J.~P. Anderson\inst{6}
\and L. Boldt\inst{7}
\and A. Campillay\inst{3}
\and S. Castellon\inst{3}
\and C. Contreras\inst{3}
\and G. Folatelli\inst{8}
\and W.~L. Freedman\inst{4}
\and M. Hamuy\inst{6}
\and W. Krzeminski\inst{9}
\and G. Leloudas\inst{10,11}
\and K. Maeda\inst{8}
\and S.~E. Persson\inst{4}
\and M. Roth\inst{3}
\and N.~B. Suntzeff\inst{12,13}}

\institute{The Oskar Klein Centre, Department of Astronomy, Stockholm University, AlbaNova, 10691 Stockholm, Sweden
\and
 Department of Physics and Astronomy, Aarhus University, Ny Munkegade 120, DK-8000 Aarhus C, Denmark
 \and
Carnegie Observatories, Las Campanas Observatory, 
  Casilla 601, La Serena, Chile
\and
  Observatories of the Carnegie Institution for Science, 813 Santa Barbara Street, Pasadena, CA, USA
  \and
INAF-Osservatorio Astronomico di Padova, vicolo dell'Osservatorio 5, 35122 Padova, Italy
  \and
Departamento de Astronomia, Universidad de Chile, Casilla 36D, Santiago, Chile
\and
Argelander Institut f\"ur Astronomie, Universit\"at Bonn, Auf dem H\"ugel 71, D-53111 Bonn, Germany
\and
Kavli Institute for the Physics and Mathematics of the Universe (IPMU), University of Tokyo, 5-1-5 
Kashiwanoha, Kashiwa, Chiba 277-8583, Japan
\and
N. Copernicus Astronomical Center, ul. Bartycka 18, 00-716 Warszawa,
Poland
\and
The Oskar Klein Centre, Department of Physics, Stockholm University, AlbaNova, 10691 Stockholm, Sweden
\and
Dark Cosmology Centre, Niels Bohr Institute, University of Copenhagen, 2100 Copenhagen, Denmark
\and
Department of Physics and Astronomy, Texas A\&M University, College Station, TX 77845, USA
\and
The Mitchell Institute for Fundamental Physics and Astronomy, Texas A\&M University, College Station, TX 77845, USA
}

\date{Received 26 July 2012 / Accepted 21 August 2012}

\abstract
{}
{We provide additional observational evidence that some Type Ia supernovae (SNe~Ia) show signatures of circumstellar interaction (CSI) with hydrogen-rich material.}
{Early phase optical and near-infrared (NIR) light curves and spectroscopy of SN~2008J obtained 
by the {{\em Carnegie Supernova Project}} are studied and compared to those of SNe~2002ic and 2005gj. Our NIR spectrum is the first obtained for a 2002ic-like object extending up to 2.2~$\mu$m. A published high-resolution spectrum is used to provide insight on the circumstellar material (CSM).}
{SN~2008J is found to be affected by $A_V\sim$~1.9 mag of extinction and to closely resemble SN~2002ic. Spectral and color comparison to SNe~2002ic and 2005gj suggests $R_V$ $<$ 3.1. Spectral decomposition reveals the underlying SN emission matches a 1991T-like event and, since SN~2008J is as luminous as SN~2005gj ($V_{\rm max}$ $=$ $-$20.3 mag), we conclude that their CSI emissions are similarly robust. The high-resolution spectrum reveals narrow emission lines produced from un-shocked gas characterized by a wind velocity of $\sim$ 50 km~s$^{-1}$. We conclude that SN~2008J best matches an explosion of a SN~Ia that interacts with its CSM.}
{}


\maketitle
\section{Introduction}
\label{sec:intro}
Supernova (SN)~2002ic was the first event identified as a Type Ia SN interacting with hydrogen-rich circumstellar material (CSM) \citep{hamuy03,wood04}.
Spectroscopically similar to the bright SN~1991T, the 
 spectral energy distribution of SN~2002ic also exhibited
prevalent narrow Balmer emission lines which are typically produced by SN--CSM interaction (CSI) in Type IIn core-collapse (CC) SNe. 
In SN~2002ic, the CSI explains not only its high peak bolometric luminosity
(L$_{\rm bol}$ $\approx$ 3$\times$10$^{43}$~erg~s$^{-1}$) and slow declining light curve, but also its strong, broad hydrogen, calcium, and iron features observed at late epochs \citep{chugai04}.
SN~2002ic-like events are extremely rare. 
To date only SN~1997cy \citep{germany00,turatto00,hamuy03,deng04}, SN~1999E 
\citep{rigon03}, SN~2005gj \citep{aldering06,prieto07} and PTF11kx \citep{dilday12} have been found to resemble SN~2002ic. Given the rarity of this kind of transient, and the opportunity they offer to better understand the progenitors of SNe~Ia, it is imperative 
to enlarge the observational sample of 2002ic-like SNe. 
 This letter presents early-time optical and near-infrared (NIR) 
observations of the 2002ic-like SN~2008J obtained by the 
{\em Carnegie Supernova Project} (CSP; \citealp{hamuy06}).

\section{Observations}
\label{sec:obs}
SN~2008J was discovered near the center of the SBbc galaxy 
MGC$-$02$-$07$-$033 on 15.2 January 2006 UT \citep{thrasher08}, 
and two days later was classified as a Type-Ia/IIn SN \citep{stritzinger08}.
According to \citet{ned} (via NED), the redshift of the host is $z = 0.0159$, which corresponds to a luminosity distance of $66.3\pm1.2$~Mpc, when we adopt WMAP5 cosmology and corrections for peculiar motions.
The CSP began photometric follow-up 4 days after discovery and continued over the course of 34 days. 
In doing so, 10 epochs of optical and 6 epochs of NIR imaging 
were obtained with the 1-m Swope telescope at Las Campanas Observatory.
All science images were reduced in a standard manner following methods described in \citet{contreras10}.
Photometry of the SN was computed differentially with respect to local sequence stars.
 Optical and NIR local sequences in the standard photometric systems are listed in Table~\ref{tab:localsequences}, and optical and NIR photometry in the {\em natural system} of the Swope telescope is presented in Tables~\ref{tabphot} and \ref{tabphotIR}.
Fig.~\ref{absmag} displays the absolute magnitude optical and NIR light
 curves of SN~2008J, compared to those of SNe~2002ic and 2005gj.   
The CSP obtained three optical spectra of SN~2008J, which were reduced following the method described by \citet{hamuy06}. 
   A high-resolution optical spectrum was acquired by \citet{sternberg11}.
    This spectrum is used to place constraints
   on the nature of the CSM.
A NIR spectrum of SN~2008J was also obtained with the NTT 5 days before $B$-band maximum ($B_{\rm max}$). This is the first published NIR spectrum of a 2002ic-like object extending up to 2.2~$\mu$m. A log of these spectral observations is given in Table~\ref{tabspectra} and the data are plotted in Figs.~\ref{spec_comp}--\ref{spec_fit}.

\section{Results}
\label{sec:analysis}

\begin{figure*}[t]
\centering
\includegraphics[width=7.4in]{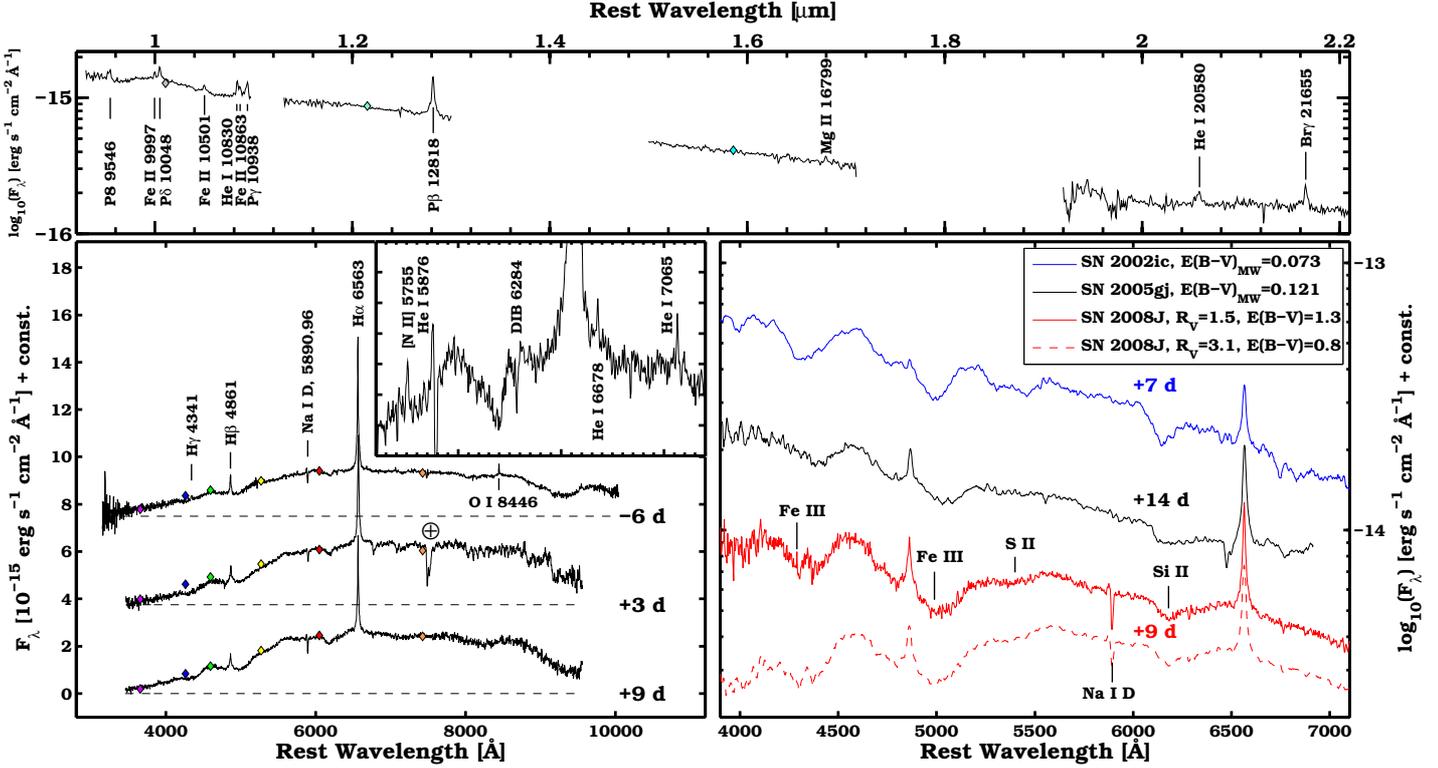}
\caption{{\em (Top panel)} NIR spectrum of SN~2008J at 5 days before $B_{max}$. The spectrum has been scaled to match NIR  photometry (colored diamonds).
Paschen emission lines as well as \ion{He}{i} and \ion{Fe}{ii} are visible.
{\em (Left panel)} Optical spectroscopy of SN~2008J. Each spectrum has been scaled to match optical photometry. Days since $B_{max}$
are reported, and horizontal dashed lines correspond to zero flux level.
 Prominent Balmer lines, \ion{Na}{i}~D, narrow \ion{O}{i} $\lambda$8446 and telluric features are labelled. In the inset we indicate \ion{He}{i} narrow emission lines in the first spectrum. Features attributed to [\ion{N}{ii}] $\lambda$5755 and diffuse interstellar bands ($\lambda$6284) are also discernible. {\em (Right panel)} Spectral comparison of SN~2008J to SNe 2002ic \citep{hamuy03} and 2005gj \citep{prieto07} at comparable epochs. 
Each spectrum has been de-reddened (see legend).
 \label{spec_comp}} 
\end{figure*}

The spectral sequence reveals prevalent \ion{Na}{i}~D absorption and a
 diffuse interstellar band (DIB, $\lambda$6284) at the SN rest frame (see Fig.~\ref{spec_comp}, left panel inset), which are indicative of
significant host dust extinction (Galactic extinction is only $A_V$ $=$ 0.062~mag, \citealp{red}, via NED). Significant reddening is also consistent
 with the heavily suppressed blue-end of the optical spectrum, and with the mid-IR dust emission documented by \citet{fox11}.
We turn to the
high-resolution spectrum
 shown in Fig.~\ref{spec_fit} (top-left panel), in order to fit Gaussians to the individual \ion{Na}{i}~D components. 
 We obtain total equivalent widths (EWs) of
\ion{Na}{i}~D2 $=$ 1.87~\AA\ and \ion{Na}{i}~D1 $=$ 1.47~\AA. These measurements indicate that the absorption features are saturated and that $E(B-V)_{\rm host}$ exceeds 0.5~mag (\citealp{turatto03}, although see \citealp{poznanski11}) for $R_V$ $=$ 3.1 \citep{cardelli89}.

The first optical and the NIR spectra are used to estimate an upper limit on $E(B-V)_{\rm tot}$, 
through the Balmer decrement and the comparison of P$\beta$ to H$\alpha$ and H$\beta$ ratios to
 theoretical expectations. This limit is $R_V$--dependent and we note that several studies (e.g. \citealp{wang08,folatelli10}) revealed that $R_V$ values for SNe~Ia may
 differ from that of normal interstellar dust, pointing to $R_V$ as low as 1.5.
We find $E(B-V)_{\rm tot}$ $\le$ 0.8$\pm$0.2~mag and 1.3$\pm$0.3~mag for $R_V$ $=$ 3.1 and 1.5
 respectively. 

\begin{figure}[t]
\centering
\includegraphics[width=3.6in]{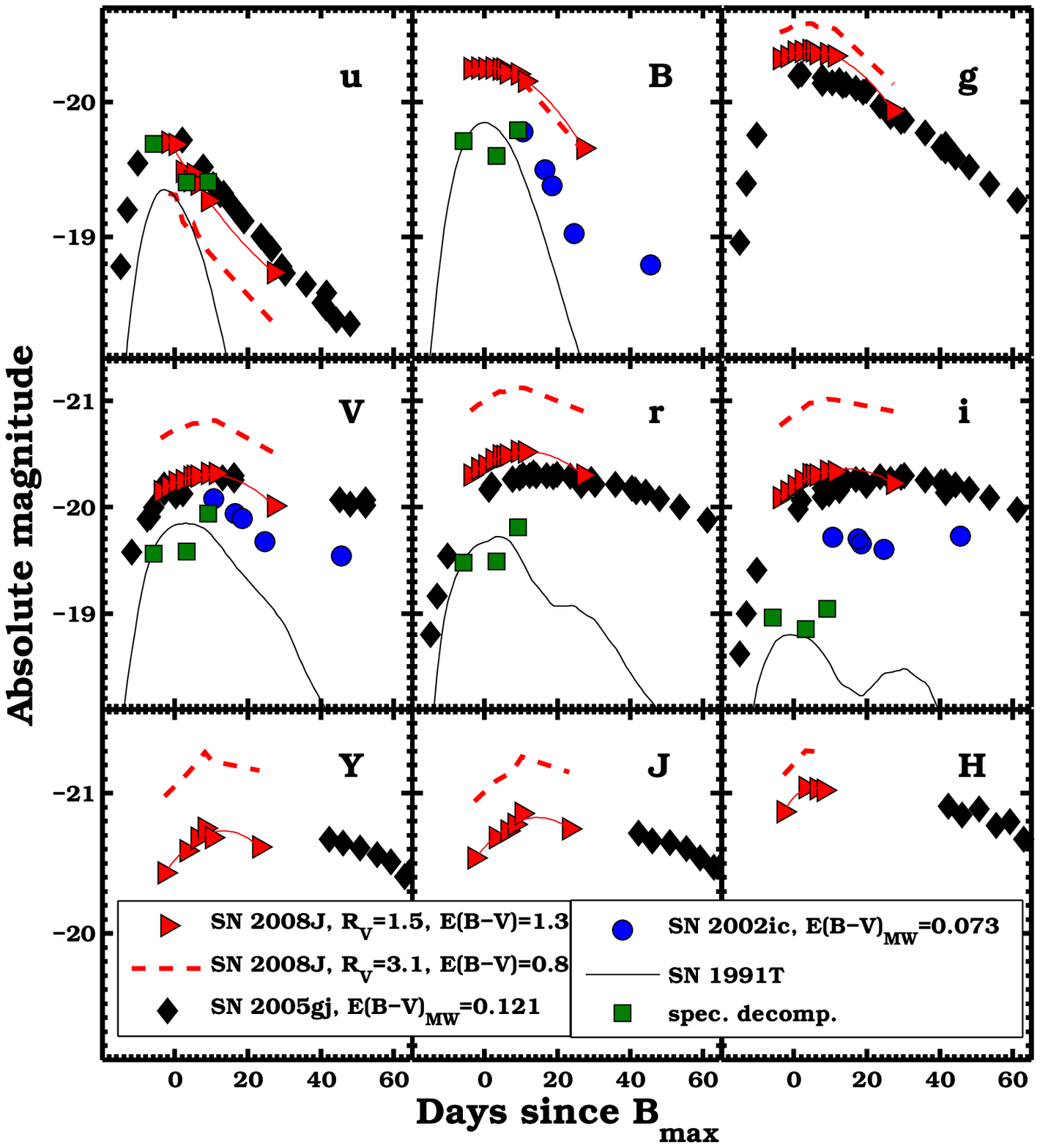}
\caption[]{Absolute magnitude light curves of SN~2008J corrected
for different amounts of reddening as discussed in the text.
For comparison de-reddened absolute magnitudes of SNe~2002ic \citep{hamuy03} and 2005gj \citep{aldering06,prieto07} are also plotted, assuming distance moduli of 37.32~mag and 37.09~mag  respectively.
Template optical light curves of SN~1991T \citep{nugent02} are plotted and scaled
to match the absolute $uBVri$ peak magnitudes of SN~1991T presented by \citet{prieto07} and \citet{saha01}. Synthetic magnitudes from the spectral decompostion of SN~2008J (see also Fig.~\ref{decomp}) match reasonably well the SN~1991T templates.
 \label{absmag}}
\end{figure}

In Fig.~\ref{absmag} we compare the absolute magnitude light curves of SN~2008J, corrected for these two different reddenings, with those of SNe~2002ic and 2005gj.  Peak apparent and absolute magnitudes of SN~2008J are listed in Table~\ref{tabmaxx}. SN~2008J peaked first in the $u$- and $B$-bands, and subsequently at longer wavelengths.
For $R_V$ $=$ 1.5 coupled 
 with $E(B-V)_{\rm tot}$ $=$ 1.3~mag, SN~2008J appears to be as luminous as SN~2005gj in each of the passbands.
Instead, for a $R_V$ $=$ 3.1 and $E(B-V)_{\rm tot}$ $=$ 0.8~mag, peak magnitudes would be significantly brighter (0.5--0.7~mag) in the $Vri$-bands whereas the $u$-band peak would be fainter (0.4~mag).
Therefore, using the standard $R_V$ $=$ 3.1 would make SN~2008J much redder than the other SNe, even for the upper limit $E(B-V)_{\rm tot}$ $=$ 0.8~mag. Instead, a lower $R_V$ gives rise to colors similar to those of SNe~2002ic and 2005gj.
 
This is confirmed in Fig.~\ref{spec_comp} (right panel), where the last (de-reddened) spectrum of SN~2008J is plotted along with similar phase spectra of SNe~2002ic and 2005gj. The comparison reveals that SN~2008J strongly resembles both SNe~2002ic and 2005gj and that its colors are similar to those of the other two SNe for $R_V$ $=$ 1.5 and $E(B-V)_{\rm tot}$ $=$ 1.3~mag. Therefore, assuming that 2002ic-like events have similar colors, we adopt $R_V$ $=$ 1.5 and $E(B-V)_{\rm tot}$ $=$ 1.3$_{-0.2}^{+0.3}$~mag. Here the uncertainty of $-0.2$~mag is obtained from the light curve comparison, which reveals (for the adopted $R_V$ $=$ 1.5)
that a color excess lower than 1.1~mag would make the $u$- and $B$-band peak
 magnitudes of SN~2008J fainter than those of a 1991T-like event. These assumptions yield a visual extinction of A$_V$ $=$ 1.93$_{-0.30}^{+0.45}$~mag.

The low-resolution optical spectra of SN~2008J are characterized by a red continuum with prevalent H$\alpha$ and 
H$\beta$ emission lines superposed. The continuum also reveals broad features which become 
more prominent as the SN is expanding, e.g. the structures with rounded peaks at 
$\sim$4700, 5600 and 8700~\AA\ (see Fig.~\ref{spec_comp}, left panel). 


After continuum subtraction, the Balmer lines are well represented by the sum of two Lorentzian profiles having the same peak wavelength (see Fig.~\ref{spec_fit}, bottom-left panels). H$\alpha$ and H$\beta$ are characterized by a narrow, unresolved component ($v_{n}$~$\lesssim$~500~km~s$^{-1}$) on top of a broader, resolved component characterized by $v_{b}$~$\approx$ 1500~km~s$^{-1}$.
Lines having only narrow, unresolved component 
characterize the first spectrum of SN~2008J. In the inset of 
Fig.~\ref{spec_comp} we highlight the presence
of \ion{He}{i}~$\lambda\lambda$7065, 6678, 5876 and \ion{N}{ii}~$\lambda$5755 features. 
\ion{O}{i}~$\lambda$8446 is also identified in the first spectrum shown in the left 
panel of Fig~\ref{spec_comp}. Plotted in  Fig.~\ref{spec_fit} (right panels) are 6 narrow lines from the high-resolution ($\sim$5--8~km~s$^{-1}$) spectrum, 4 of them exhibiting P-Cygni profiles.\footnote{H$\alpha$ is missing as it falls between two orders of the echelle spectrum.} 
These features suggest the presence of slowly moving material surrounding the SN, and they exclude the possibility that they are associated with an underlying \ion{H}{ii} region.
Fitting each line with a Lorentzian profile provides a measure of the FWHM, 
which are summarized in Table~\ref{fwhm}, and indicates that the CSM has a 
velocity of 52$\pm$15~km~s$^{-1}$. 
The high resolution spectrum also exhibits DIBs at $\lambda_0$ 6284 and 5780~\AA, having EW $=$ 0.6~\AA\ and 0.2~\AA\ respectively.

The broad features in the low-resolution optical spectra are identified through the comparison with SNe~2002ic and 2005gj.
The spectra show a dip around $\lambda_{0}$ 6180~\AA, which likely originates from \ion{Si}{ii}~$\lambda$6355 blue-shifted by $\sim$8200~km~s$^{-1}$. 
\ion{Fe}{iii}~$\lambda$5129 is probably the line producing the dip at $\sim$4990~\AA,
whereas \ion{Fe}{iii}~$\lambda$4404 absorption is visible at $\sim$4290~\AA. \ion{S}{ii}~$\lambda$5468 and \ion{S}{ii}~$\lambda$5633 might cause the depression at $\sim$5400~\AA.
The NIR spectrum exhibits narrow, unresolved emission lines, including P$\beta$, P$\delta$, P$8$, Br$\gamma$, \ion{He}{i} $\lambda\lambda$10830, 20580 and \ion{Fe}{ii} $\lambda\lambda$9997, 10501, 10863. The bright NIR hydrogen lines (P$\beta$, Br$\gamma$) are also characterized by a broad component, similar to that observed for the Balmer lines.

\section{Discussion}
\label{sec:discussion}
\begin{figure}[h]
\centering
\includegraphics[width=3.6in]{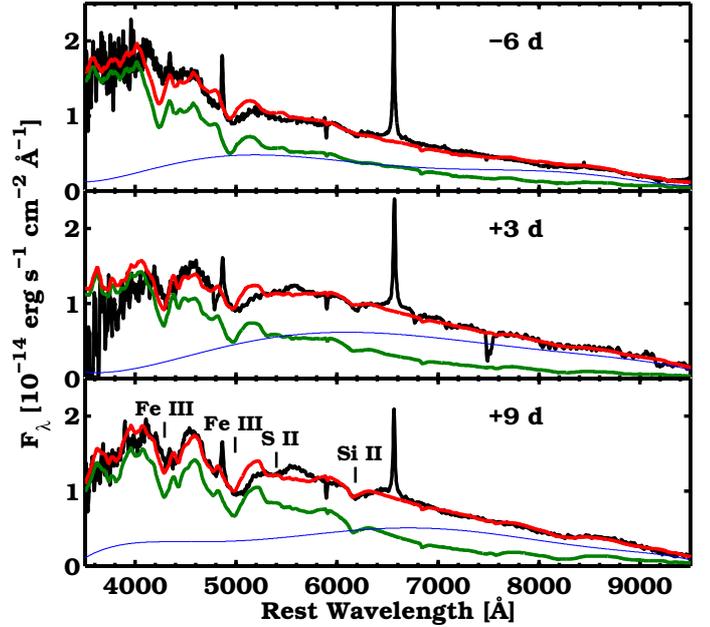}
\caption{Spectral decomposition of SN~2008J. The de-reddened spectra are well reproduced by the sum (red) of a smooth continuum (blue), powered by the CSI, and of a scaled SN~1991T template spectrum (green). We used SN~1991T templates at the same phase (measured since $B_{\rm max}$) of the spectra of SN~2008J, which corresponds to 17, 26 and 32 days since explosion. \label{decomp}} 
\end{figure}

\citet{hamuy03} and \citet{aldering06} interpreted SNe~2002ic and 2005gj to
be SNe~Ia interacting with H-rich CSM. This was 
based mainly on the fact that their
spectra are well represented by the sum of two components: a smoothly varying continuum (produced by the CSI) and a diluted spectrum of a 1991T-like SN at comparable phase.
In Fig.~\ref{decomp} we show a similar decomposition for the 3 optical spectra of SN~2008J, where we used low-order polynomials to fit the continuum.
The models match the spectra reasonably well, especially the \ion{Si}{ii} dip at $\sim$6200 \AA, and the features between $\sim$3500--5000~\AA. However, the rounded peak at $\sim$5600 \AA\ is not perfectly reproduced, although a faint feature associated with \ion{S}{ii} is also present at the same wavelength in the spectrum of SN~1991T. 
As in SN~2005gj, the notch characterizing the spectrum of SN~1991T at 5300~\AA\ is barely detected in SN~2008J. The imperfect match is not surprising since the spectral decomposition model does not include the coupling between CSI radiation and
SN envelope, which should alter the spectral features of the underlying SN.
As SN~2008J appears to be as luminous as SN~2005gj (see Fig.~\ref{absmag}), we conclude that the emission due to the CSI is also similar. The flux contribution of the underlying SN to the $V$-band maximum of both SNe is $\sim$60\%.
Similarly to SNe~2002ic and 2005gj, if the CSM is optically thick, the interaction region should not completely cover the SN~Ia since its broad features are evident in the spectra.
We note that \citet{benetti06} argued that SN~2002ic might be better explained by a Type Ic event like SN~2004aw \citep{taubenberger06}, which also shows the presence of \ion{Si}{II}, rather than by the CSM-SN~Ia scenario suggested by \citet{hamuy03}.
 However, in the recent case of PTF11kx, the first spectrum obtained unequivocally shows a 1991T-like SN whose spectrum was only barely affected by CSI \citep{dilday12}. The discovery of PTF11kx therefore provides strong support to the interpretation of the other members of the SN~2002ic class as CSM--SNe Ia. 
 
The narrow emission lines and P-Cygni profiles reveal the presence of un-shocked, radially expanding CSM photo-ionized by CSI radiation. The CSM likely originates from winds in the progenitor system, and it is characterized by a velocity $v_{w}\approx$~50~ km~s$^{-1}$.
This velocity is higher than that of red supergiant (RSG) or asymptotic giant branch (AGB) winds ($\sim$10~km~s$^{-1}$), and much lower than that of Wolf-Rayet (WR) winds (up to 2000~km~s$^{-1}$). Post-AGB stars show wind velocities in the range of 100$-$400~km~s$^{-1}$ \citep{kotak04}.
 However, we can not exclude the possibility that the precursor wind was accelerated by photoionization heating as noticed by \citet{aldering06} for SN~2005gj. This would make the AGB or RSG wind velocities more compatible with the measured ones.
Episodic nova events (like that suggested for PTF11kx) can also give rise to CSM velocities of 50$-$100~km~s$^{-1}$ \citep{dilday12}. However, PTF11kx shows a complex CSM, with multiple wind velocities which we do not find in SN~2008J.

The broad components that we observe in bright hydrogen lines are probably related to the shock region. 
The absolute values of the Balmer decrement are higher for the narrow 
component than for the broad one, indicating that they 
are produced in different regions. 
When the CSI dominates the emission, the luminosity of the 
broad component of H$\alpha$, $L$(H$\alpha_{\rm broad}$), is  proportional to 
the kinetic energy dissipated
per unit time across the shock front \citep{salamanca98}.
Therefore, we roughly estimate the mass-loss rate to be 
$\dot{M}\approx3\times10^{-3}~M_{\odot}~$yr$^{-1}$. 
This is slightly lower than the mass-loss rate computed for SN~2002ic 
\citep{kotak04}.
Here we have adopted an efficiency factor $\epsilon_{\rm H{\alpha}}=0.1$, shock and wind 
velocities as measured from the full-width-at-zero-intensity (FWZI) of the broad and narrow H$\alpha$ components (6000 and 100~km~s$^{-1}$ respectively), and a measured $L$(H$\alpha_{\rm broad}$) $=$  1.3$\times$10$^{41}$~erg~s$^{-1}$.

The host galaxy of SN~2008J is a bright spiral galaxy having $M_{B}=-$20.2~mag \citep{doyle05}. Therefore SN~2008J, like PTF11kx and SN~1999E, is not located in an unusual environment as in the case of SNe~1997cy, 2002ic and 2005gj, all of which occurred in faint hosts. The significant amount of MIR-emission found by \citet{fox11} at late times (593 days since discovery)
 indicates the presence of $\sim$0.01~$M_{\odot}$ of circumstellar dust.
 Late time (+380 days) dust emission was also observed in SN~2002ic \citep{kotak04}. We conclude that SN~2008J resulted from CSI of a 1991T-like event, similar to SNe~2002ic, 2005gj and PTF11kx. The observations of these objects suggest the idea that more efficient CSI may occur leading to the misidentification of interacting SNe~Ia as SNe~IIn.

\begin{acknowledgements}

We thank J.~L. Prieto for sharing spectra of SN~2005gj with us.
This research has made use of the Keck Observatory Archive (KOA). 
 This material is based upon work supported by NSF under 
grants AST--0306969, AST--0607438, and AST--1008343. 
The Oskar Klein Centre is funded by the Swedish Research Council. 
J.~P. A. and M. H. acknowledge support by
CONICYT through FONDECYT grant 3110142, and by the Millennium Center for
Supernova Science (P10-064-F).
N.~B. S. acknowledges support from the Mitchell Institute for Fundamental Physics and Astronomy,  NSF grant AST--1008343 and the Mitchell Chair in Observational Astronomy.
G.~L. is supported by the Swedish Research Council through grant 623--2011--7117.
DARK is funded by the DNRF.
M.~D. S., G. F. and K. M. acknowledge the support by World Premier International Research Center Initiative, MEXT, Japan.
\end{acknowledgements}

\bibliographystyle{aa}

\onecolumn
\onlfig{4}{
\begin{figure}[h]
\centering
\includegraphics[width=7.4in]{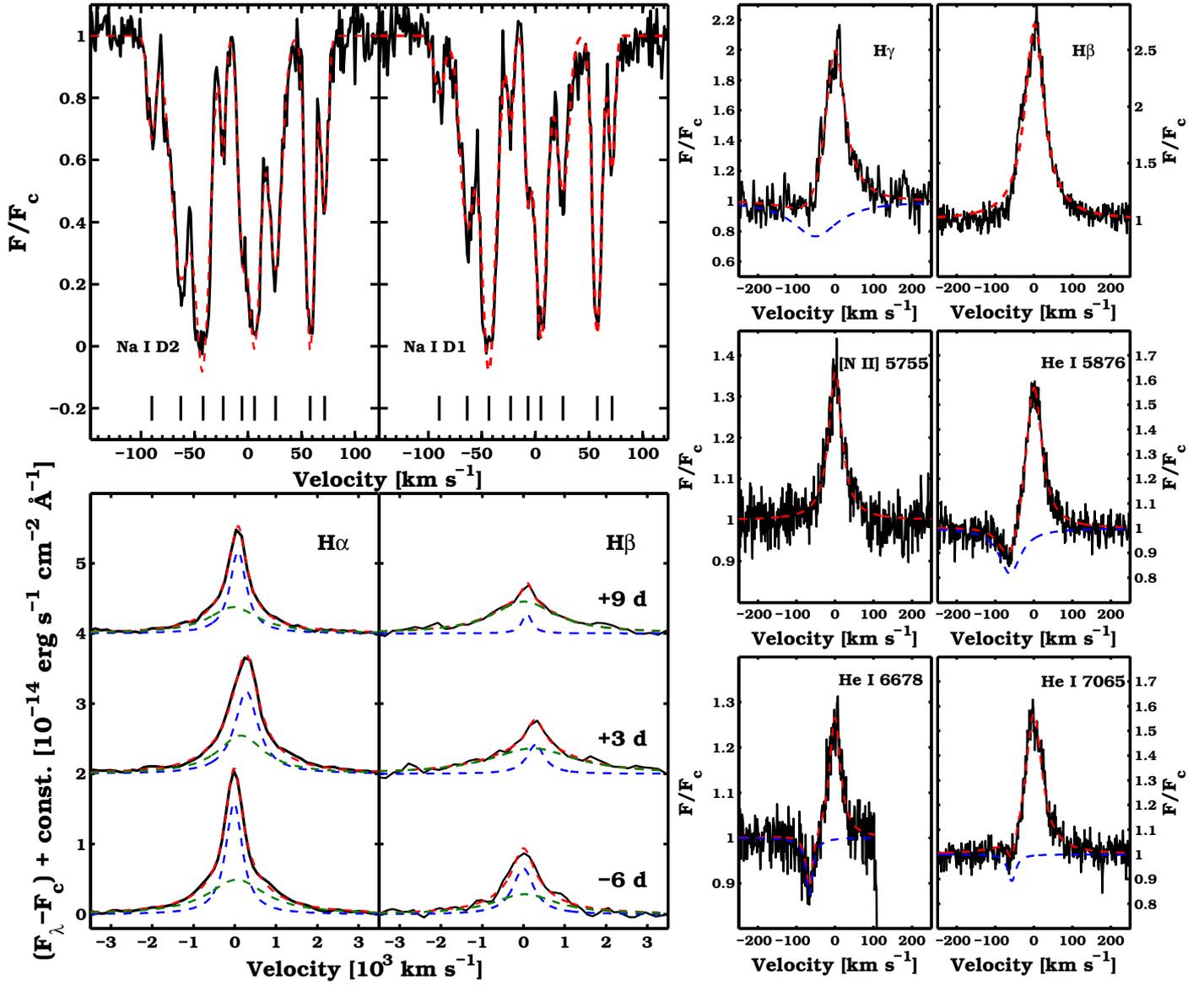}
\caption{{\em (Top-left panels)} 
\ion{Na}{i}~D absorption features in the high resolution spectrum of SN~2008J. Several components are present and their peak positions are indicated by black solid lines. 
The features are fit with the sum of 9 Gaussian functions, the best fit is shown in red.
Velocity is plotted in the SN rest frame, as determined from the narrow emission line peaks in the same high resolution spectrum {\em (Bottom-left panels)}
H$\alpha$ and H$\beta$ sequences from the low resolution spectra, shown after low-order polynomial continuum subtraction. Each profile is
well fit by the sum (in red) of two Lorentzian components (in blue and green).
{\em (Right panels)} 
Narrow emission lines in the high-resolution spectrum of SN~2008J. Each line has been fit with a Lorentzian profile for the emission and one for the absorption when detected. The absorption component is shown in blue, the total best fit in red. The spectrum is presented in velocity scale, where the zero velocity corresponds to the emission peak position.
 \label{spec_fit}} 
\end{figure}}

\clearpage
\onltab{1}{
\begin{deluxetable} {lllcccccccccc}
\rotate
\tabletypesize{\tiny}
\tablecolumns{13}
\tablewidth{0pt}
\tablecaption{Optical and near-infrared photometry of the local sequences in the standard system\label{tab:localsequences}}
\tablehead{
\colhead{} &
\colhead{}  &
\colhead{}  &
\colhead{$u'$}   &
\colhead{$g'$}   &
\colhead{$r'$}   &
\colhead{$i'$}   &
\colhead{$B$}   &
\colhead{$V$}  &
\colhead{$Y$}   &
\colhead{$J$}   &
\colhead{$H$}  &
\colhead{$K_s$}\\
\colhead{STAR} &
\colhead{$\alpha~(2000)$}  &
\colhead{$\delta~(2000)$}  &
\colhead{(mag)}   &
\colhead{(mag)}   &
\colhead{(mag)}   &
\colhead{(mag)}   &
\colhead{(mag)}   &
\colhead{(mag)}  &
\colhead{(mag)}   &
\colhead{(mag)}   &
\colhead{(mag)}  &
\colhead{(mag)}}
\startdata
01      & 02:34:42.34   & $-$10:50:57.55        & $ \cdots $    & $ \cdots $    & $ \cdots $    & $ \cdots $    & $ \cdots $    & $ \cdots $    &  11.387(020)  &  11.180(014)  & $ \cdots $    & $ \cdots $    \\
02      & 02:34:32.13   & $-$10:54:08.89        &  17.014(032)  &  15.323(008)  &  14.743(007)  &  14.517(008)  &  15.767(014)  &  14.979(008)  &  13.778(020)  &  13.540(020)  & $ \cdots $    & $ \cdots $    \\
03      & 02:34:21.64   & $-$10:53:03.59        &  17.948(024)  &  15.706(010)  &  14.904(007)  &  14.629(008)  &  16.230(020)  &  15.263(008)  &  13.860(017)  &  13.544(014)  & $ \cdots $    & $ \cdots $    \\
04      & 02:34:29.09   & $-$10:54:11.99        &  18.864(113)  &  16.284(010)  &  14.999(011)  &  14.373(012)  &  16.974(017)  &  15.591(010)  &  13.295(020)  &  12.910(020)  & $ \cdots $    & $ \cdots $    \\
05      & 02:34:20.76   & $-$10:54:10.12        &  17.203(033)  &  16.091(008)  &  15.681(007)  &  15.523(008)  &  16.427(016)  &  15.842(008)  &  14.885(020)  &  14.667(020)  & $ \cdots $    & $ \cdots $    \\
06      & 02:34:12.06   & $-$10:48:14.11        &  17.813(016)  &  16.282(007)  &  15.713(007)  &  15.506(008)  &  16.719(016)  &  15.952(008)  & $ \cdots $    & $ \cdots $    & $ \cdots $    & $ \cdots $    \\
07      & 02:34:21.89   & $-$10:49:23.30        &  19.563(077)  &  17.130(010)  &  16.179(007)  &  15.827(009)  &  17.686(017)  &  16.623(008)  &  14.995(026)  &  14.559(072)  &  13.934(053)  &  13.884(014)  \\
08      & 02:34:22.98   & $-$10:52:35.72        &  18.622(054)  &  17.143(012)  &  16.583(007)  &  16.364(008)  &  17.572(012)  &  16.817(008)  &  15.677(017)  &  15.415(014)  & $ \cdots $    & $ \cdots $    \\
09      & 02:34:22.67   & $-$10:52:19.16        &  20.497(135)  &  17.842(020)  &  16.526(007)  &  15.899(008)  &  18.555(011)  &  17.141(008)  &  14.886(014)  &  14.472(014)  & $ \cdots $    & $ \cdots $    \\
10      & 02:34:33.03   & $-$10:51:38.70        &  18.456(035)  &  17.529(009)  &  17.180(007)  &  17.051(008)  &  17.814(016)  &  17.317(008)  &  16.456(069)  &  16.270(020)  & $ \cdots $    & $ \cdots $    \\
11      & 02:34:37.60   & $-$10:51:12.64        &  20.776(166)  &  18.090(017)  &  16.888(008)  &  16.330(008)  &  18.748(015)  &  17.455(009)  &  15.361(028)  &  14.941(014)  & $ \cdots $    & $ \cdots $    \\
12      & 02:34:38.22   & $-$10:52:13.37        &  18.861(024)  &  17.959(009)  &  17.576(007)  &  17.422(008)  &  18.300(018)  &  17.728(008)  &  16.880(024)  &  16.664(020)  & $ \cdots $    & $ \cdots $    \\
13      & 02:34:09.23   & $-$10:52:55.63        &  20.654(160)  &  18.193(010)  &  17.122(007)  &  16.702(008)  &  18.802(023)  &  17.624(008)  & $ \cdots $    & $ \cdots $    & $ \cdots $    & $ \cdots $    \\
14      & 02:34:13.94   & $-$10:49:16.61        &  19.598(076)  &  19.070(034)  &  18.926(027)  &  18.916(034)  &  19.306(064)  &  19.001(038)  & $ \cdots $    &  18.004(085)  & $ \cdots $    & $ \cdots $    \\
15      & 02:34:21.44   & $-$10:47:44.38        & $ \cdots $    &  19.517(021)  &  18.174(008)  &  17.245(015)  &  20.342(043)  &  18.755(020)  &  16.029(020)  &  15.635(020)  & $ \cdots $    & $ \cdots $    \\
16      & 02:34:32.12   & $-$10:48:58.86        & $ \cdots $    &  19.837(020)  &  18.435(026)  &  16.847(023)  &  20.711(060)  &  19.052(020)  &  15.148(031)  &  14.655(014)  & $ \cdots $    & $ \cdots $    \\
17      & 02:34:16.79   & $-$10:48:28.44        & $ \cdots $    & $ \cdots $    & $ \cdots $    & $ \cdots $    & $ \cdots $    & $ \cdots $    &  16.721(017)  &  16.251(014)  & $ \cdots $    & $ \cdots $    \\
18      & 02:34:29.84   & $-$10:52:05.95        & $ \cdots $    & $ \cdots $    & $ \cdots $    & $ \cdots $    & $ \cdots $    & $ \cdots $    &  17.243(056)  &  16.815(024)  & $ \cdots $    & $ \cdots $    \\
19      & 02:34:22.32   & $-$10:52:44.90        & $ \cdots $    & $ \cdots $    & $ \cdots $    & $ \cdots $    & $ \cdots $    & $ \cdots $    &  17.169(035)  &  16.920(052)  & $ \cdots $    & $ \cdots $    \\
20      & 02:34:20.29   & $-$10:51:52.74        & $ \cdots $    & $ \cdots $    & $ \cdots $    & $ \cdots $    & $ \cdots $    & $ \cdots $    &  17.283(029)  &  16.895(099)  & $ \cdots $    & $ \cdots $    \\
21      & 02:34:23.43   & $-$10:49:01.24        & $ \cdots $    & $ \cdots $    & $ \cdots $    & $ \cdots $    & $ \cdots $    & $ \cdots $    &  17.813(081)  &  17.448(090)  &  16.873(145)  & $ \cdots $    \\
22      & 02:34:20.91   & $-$10:52:20.60        & $ \cdots $    & $ \cdots $    & $ \cdots $    & $ \cdots $    & $ \cdots $    & $ \cdots $    &  17.917(063)  &  17.486(063)  & $ \cdots $    & $ \cdots $    \\
23      & 02:34:29.18   & $-$10:51:05.65        & $ \cdots $    & $ \cdots $    & $ \cdots $    & $ \cdots $    & $ \cdots $    & $ \cdots $    &  18.096(037)  &  17.647(106)  &  16.812(276)  &  16.825(298)  \\
24      & 02:34:36.21   & $-$10:51:16.02        & $ \cdots $    & $ \cdots $    & $ \cdots $    & $ \cdots $    & $ \cdots $    & $ \cdots $    &  18.238(094)  &  17.974(099)  & $ \cdots $    & $ \cdots $    \\
25      & 02:34:15.59   & $-$10:49:28.42        & $ \cdots $    & $ \cdots $    & $ \cdots $    & $ \cdots $    & $ \cdots $    & $ \cdots $    &  18.354(138)  &  18.428(100)  & $ \cdots $    & $ \cdots $    \\
26      & 02:34:19.40   & $-$10:49:06.60        & $ \cdots $    & $ \cdots $    & $ \cdots $    & $ \cdots $    & $ \cdots $    & $ \cdots $    &  18.397(085)  &  18.097(120)  &  17.518(273)  & $ \cdots $    \\
27      & 02:34:32.14   & $-$10:51:48.89        & $ \cdots $    & $ \cdots $    & $ \cdots $    & $ \cdots $    & $ \cdots $    & $ \cdots $    &  16.058(080)  &  15.608(028)  & $ \cdots $    & $ \cdots $    \\
28      & 02:34:34.53   & $-$10:48:45.14        & $ \cdots $    & $ \cdots $    & $ \cdots $    & $ \cdots $    & $ \cdots $    & $ \cdots $    & $ \cdots $    &  18.190(081)  & $ \cdots $    & $ \cdots $    \\
29      & 02:34:17.50   & $-$10:49:57.54        & $ \cdots $    & $ \cdots $    & $ \cdots $    & $ \cdots $    & $ \cdots $    & $ \cdots $    &  17.694(052)  &  17.245(079)  & $ \cdots $    & $ \cdots $    \\
30      & 02:34:16.18   & $-$10:49:55.88        & $ \cdots $    & $ \cdots $    & $ \cdots $    & $ \cdots $    & $ \cdots $    & $ \cdots $    &  18.114(256)  &  17.686(049)  & $ \cdots $    & $ \cdots $    \\
31      & 02:34:19.77   & $-$10:47:04.88        & $ \cdots $    & $ \cdots $    & $ \cdots $    & $ \cdots $    & $ \cdots $    & $ \cdots $    &  17.806(049)  &  17.604(071)  & $ \cdots $    & $ \cdots $    \\
32      & 02:34:23.86   & $-$10:46:19.13        & $ \cdots $    & $ \cdots $    & $ \cdots $    & $ \cdots $    & $ \cdots $    & $ \cdots $    &  18.206(070)  &  18.340(128)  & $ \cdots $    & $ \cdots $    \\
\enddata 
\tablecomments{Uncertainties given in parentheses in thousandths of a
  magnitude correspond to an rms of the magnitudes obtained on
  photometric nights.}
\end{deluxetable}}

\clearpage
\onltab{2}{
\begin{deluxetable}{clccccccc}
\tablewidth{0pt}
\tabletypesize{\scriptsize}
\tablecaption{Optical photometry of SN~2008J in the natural system.\label{tabphot}}
\tablehead{
\colhead{JD~-~$2,453,000$} &
\colhead{Phase\tablenotemark{a}} &
\colhead{$u$ (mag)}&
\colhead{$g$ (mag)}&
\colhead{$r$ (mag)}&
\colhead{$i$ (mag)}&
\colhead{$B$ (mag)}&
\colhead{$V$ (mag)}}
\startdata
1480.69 & $-$7.66 & \ldots &  \ldots      &   15.4\tablenotemark{b}    &\ldots &\ldots  &  \ldots               \\
1481.70 & $-$6.65 & \ldots &  \ldots      &   15.4\tablenotemark{b}    &\ldots &\ldots  &  \ldots                \\    
1484.54 & $-$3.81 & \ldots     &  16.633(009)&    15.392(005)&  14.994(004)&  17.231(012)&  16.018(008)\\ 
1486.59 & $-$1.76 & 18.609(057)&  16.616(008)&    15.333(006)&  14.936(006)&  17.225(011)&  15.972(008)\\ 
1488.56 & $+$0.21 & 18.624(051)&  16.582(006)&    15.306(005)&  14.885(005)&  17.231(013)&  15.934(007)\\ 
1490.55 & $+$2.20 & 18.824(029)&  16.579(007)&    15.249(005)&  14.835(006)&  17.224(009)&  15.917(007)\\ 
1492.58 & $+$4.23  & 18.897(042)&  16.568(007)&    15.210(006)&  14.778(008)&  17.234(010)&  15.885(008)\\ 
1493.59 & $+$5.24  & 18.845(030)&  16.573(006)&    15.215(007)&  14.780(006)&  17.235(009)&  15.876(007)\\ 
1494.54 & $+$6.19  & 18.918(030)&  16.593(009)&    15.207(006)&  14.780(008)&  17.263(009)&  15.878(007)\\ 
1497.54 & $+$9.19  & 19.044(049)&  16.595(007)&    15.173(004)&  14.746(005)&  17.271(009)&  15.849(008)\\ 
1499.55 & $+$11.20  & \ldots     &  16.611(006)&    15.177(004)&  14.751(006)&  17.325(009)&  15.851(006)\\ 
1515.52 & $+$27.17 & 19.578(124)&  17.018(018)&    15.395(005)&  14.857(005)&  17.821(017)&  16.157(008)\\  
\enddata 
\tablecomments{Values in parentheses are 1$\sigma$ measurement uncertainties in millimag. Imaging was performed with a Site3 detector attached to the Swope.}
\tablenotetext{a}{Days since $B_{max}$ (JD~2454488.35).}
\tablenotetext{b}{Unfiltered magnitudes from discovery and confirmation images \citep{thrasher08}.}
\end{deluxetable}}

\onltab{3}{
  \begin{deluxetable}{clccccc}
\tablewidth{0pt}
\tabletypesize{\scriptsize}
\tablecaption{NIR photometry of SN~2008J in the natural system.\label{tabphotIR}}
\tablehead{
\colhead{JD~-~$2,453,000$} &
\colhead{Phase\tablenotemark{a}} &
\colhead{$Y$ (mag)}&
\colhead{$J$ (mag)}&
\colhead{$H$ (mag)}&
\colhead{$K_s$ (mag)}}
\startdata
1485.60 & $-$2.75& 14.090(011) &13.877(018) &13.442(024)  &\ldots      \\
1491.55 & $+$3.20& 13.934(011) &13.734(020) &13.268(023)  &\ldots      \\
1494.59 & $+$6.24& 13.844(022) &13.688(023) &13.277(034)  &13.140(018)  \\
1496.55 & $+$8.20& 13.775(031) &13.641(044) &13.286(034)  &13.136(011)  \\
1498.58 & $+$10.23& 13.840(011) &13.561(011) &\ldots       &\ldots     \\
1511.53 & $+$23.18&13.905(043) &13.670(013) &\ldots       &\ldots       \\
1806.65 & $+$318.30&16.776(024) &\ldots      &\ldots       &\ldots     \\
\enddata 
\tablecomments{Values in parentheses are 1$\sigma$ measurement uncertainties in millimag. The imaging was performed with RetroCam attached to the Swope.}
\tablenotetext{a}{Days since $B_{max}$ (JD~2454488.35).}
\end{deluxetable}}

\onltab{4}{
\begin{deluxetable}{lclccccc}
\tabletypesize{\scriptsize}
\tablewidth{0pt}
\tablecolumns{8}
\tablecaption{Spectroscopic observations of SN~2008J.\label{tabspectra}}
\tablehead{
\colhead{Date} &
\colhead{Julian Date} &
\colhead{Phase\tablenotemark{a}} &
\colhead{Telescope} &
\colhead{Instrument} &
\colhead{Range} &
\colhead{Resolution} &
\colhead{Integration} \\ 
\colhead{(UT)} &
\colhead{JD~-~$2,453,000$} &
\colhead{(days)} &
\colhead{} &
\colhead{} &
\colhead{(\AA)} &
\colhead{(FWHM \AA)} &
\colhead{(sec)}}
\startdata
2008 Jan. 17 & 1482.57& $-$5.78   &NTT     &EMMI &   3200 -- 10200&6-9&3$\times$300\\
2008 Jan. 18 & 1483.59& $-$4.76   &NTT     &SOFI &   9400 -- 25000&20-30&1$\times$2100\\
2008 Jan. 23 & 1488.72& $+$0.37   &Keck I   & HIRES&   4110 -- 8360&0.12&2$\times$900\\
2008 Jan. 26 & 1491.57& $+$3.22   &Du~Pont &B\&C &3518 -- 9715&8&1$\times$400\\
2008 Feb. 01 & 1497.55& $+$9.20   &Du~Pont &B\&C &  3510 -- 9713&8&3$\times$500\\
\enddata
\tablenotetext{a}{Days since $B_{max}$ (JD~2454488.35).} 
\end{deluxetable}}

\onltab{5}{
\begin{deluxetable}{lcc}
\tabletypesize{\scriptsize}
\tablewidth{0pt}
\tablecaption{Parameters from the narrow emission line fit.\label{fwhm}}
\tablehead{
\colhead{Line} &
\colhead{$\lambda_c$} &
\colhead{FWHM} \\
\colhead{ } &
\colhead{(\AA)} &
\colhead{(km~s$^{-1}$)}}
\startdata
H$\gamma$~$\lambda$4341     &  4340.9    & 62$\pm$4   \\
H$\beta$~$\lambda$4861      &  4861.8    & 77$\pm$4   \\
$[$\ion{N}{ii}$]$~$\lambda$5755   &  5755.0    & 44$\pm$2   \\
\ion{He}{i}~$\lambda$5876     &  5876.2    & 51$\pm$2   \\
\ion{He}{i}~$\lambda$6678     &  6678.7    & 33$\pm$2   \\
\ion{He}{i}~$\lambda$7065     &  7065.8    & 49$\pm$1   \\
\enddata
\end{deluxetable}}

\onltab{6}{
\begin{deluxetable}{clccc}
\tabletypesize{\scriptsize}
\tablewidth{0pt}
\tablecolumns{4}
\tablecaption{Peak magnitudes of SN~2008J.\label{tabmaxx}}
\tablehead{
\colhead{Filter} &
\colhead{Phase\tablenotemark{a}} &
\colhead{Peak} &
\colhead{Peak} &
\colhead{Peak} \\ 
\colhead{} &
\colhead{(days)} &
\colhead{(apparent mag)} &
\colhead{(absolute mag)}&
\colhead{(absolute mag)}\\
\colhead{} &
\colhead{} &
\colhead{} &
\colhead{$R_V$ $=$ 1.5}&
\colhead{$R_V$ $=$ 3.1}\\
\colhead{} &
\colhead{} &
\colhead{} &
\colhead{$E(B-V)_{\rm tot}$ $=$ 1.3$_{-0.2}^{+0.3}$}&
\colhead{$E(B-V)_{\rm tot}$ $=$ 0.8$_{-0.1}^{+0.2}$}}
\startdata
$u$ &  $<$ $-$1.8       &$<$18.59        &$<$ $-$19.72$_{-0.69}^{+1.01}$   &$<$ $-$19.34$^{+1.00}_{-0.52}$  \\ 
$B$ &    0               &17.22$\pm$0.01  &$-$20.25$_{-0.56}^{+0.82}$      &$-$20.24$^{+0.88}_{-0.46}$  \\  
$g$ &  $+$4.4           &16.57$\pm$0.01  &$-$20.38$_{-0.48}^{+0.70}$       &$-$20.58$^{+0.80}_{-0.42}$ \\ 
$V$ &  $+$10.6          &15.85$\pm$0.01  &$-$20.31$_{-0.36}^{+0.52}$          &$-$20.81$^{+0.68}_{-0.36}$ \\ 
$r$ &  $+$11.5         &15.17$\pm$0.01  &$-$20.52$_{-0.28}^{+0.41}$     &$-$21.12$^{+0.59}_{-0.31}$ \\ 
$i$ &  $+$11.7         &14.75$\pm$0.01  &$-$20.34$_{-0.19}^{+0.26}$      &$-$21.04$^{+0.45}_{-0.25}$ \\ 
$Y$ &  $+$13.3          &13.79$\pm$0.01  &$-$20.73$_{-0.10}^{+0.14}$     &$-$21.24$^{+0.28}_{-0.16}$ \\ 
$J$ &  $+$14.6         &13.59$\pm$0.02  &$-$20.83$_{-0.09}^{+0.11}$      &$-$21.23$^{+0.22}_{-0.13}$ \\ 
$H$ &   $>$ $+$8.2      & $<$13.27       & $<$ $-$21.04$_{-0.07}^{+0.09}$   & $<$ $-$21.30$^{+0.16}_{-0.10}$  \\  
$K_s$&  $>$ $+$8.2      & $<$13.14       & $<$ $-$21.09$_{-0.06}^{+0.07}$  & $<$ $-$21.26$^{+0.11}_{-0.08}$ \\  
\enddata
\tablecomments{Uncertainties in the phase are $\sim$0.1 days for the optical passbands, $\sim$1 day for the NIR. Errors in the absolute magnitude are given by the uncertainties in the extinction and in the distance modulus (0.04~mag).}
\tablenotetext{a}{Days since $B_{max}$ (JD~2454488.35).} 
\end{deluxetable}}

\end{document}